\documentclass[a4paper]{jpconf}
\usepackage{graphicx}
\usepackage{amsmath, amssymb}
\usepackage{bm}
\usepackage{mathrsfs}
\usepackage{enumerate}

\newcommand{\CGN}{\mathrm{Cr_3GeN}}
\newcommand{\muSR}{\mu \mathrm{SR}}
\newcommand{\NitNMR}{$^{14}$N-NMR\ }
\newcommand{\Nitnuc}{^{14}\mathrm{N}}

\begin{document}
\title{Paramagnetic-to-nonmagnetic transition in antiperovskite nitride Cr$_3$GeN studied by $^{14}$N-NMR and $\mu$SR}

\author{K Takao$^1$, Z Liu$^1$, K Uji$^1$, T Waki$^1$, Y Tabata$^1$, I Watanabe$^2$ and H Nakamura$^1$}

\address{$^1$ Department of Materials Science and Engineering, 
 Kyoto University, Kyoto 606-8501, Japan\\
$^2$ Advanced Meson Science Laboratory, RIKEN Nishina Center, 
Wako 351-0198, Japan}

\ead{takao.kenta.42w@st.kyoto-u.ac.jp}

\begin{abstract}
The antiperovskite-related nitride Cr$_3$GeN forms a tetragonal structure with the space group $P\bar{4}2_1m$ at room temperature. It shows a tetragonal ($P\bar{4}2_1m$) to tetragonal ($I4/mcm$) structural transition with a large hysteresis at 300--400 K. 
The magnetic susceptibility of Cr$_3$GeN shows Curie-Weiss type temperature dependence at high temperature, but is almost temperature-independent below room temperature. 
We carried out $\mu$SR and \NitNMR microscopy measurements to reveal the magnetic ground state of Cr$_3$GeN. 
Gradual muon spin relaxation, which is nearly temperature-independent below room temperature, was observed, indicating that Cr$_3$GeN is magnetically inactive.
In the $^{14}$N-NMR measurement, a quadrupole-split spectrum was obtained at around $^{14}K = 0$. 
The temperature dependence of $^{14}(1/T_1)$ satisfies the Korringa relation.
These experimental results indicate that the ground state of Cr$_3$GeN is Pauli paramagnetic, without antiferromagnetic long-range order. 
\end{abstract}

\section{Introduction}
The antiperovskite compounds ($M_3M'X$, $M = $ 3d transition metal or rare earth, $M' = $ metal or metalloid, $X = $ B, C, N) have been invesestigated for several decades, and various kinds of elements with different $M$ and $M'$ have been discovered \cite{Boller1, Nardin1, Boller2, Jeitschko1, Samson1, Haschke1}.
They form a cubic structure with the space group $Pm\bar{3}m$; $M'$ atoms are located in the corner positions, while $M$ and $X$ atoms occupy the face-centered and body-centered positions, respectively.
The magnetic properties of Mn-based compounds have been widely studied with respect to their application in electrical technologies \cite{Kodama1, Kamishima1, Takenaka1}.
However, few studies have investigated the magnetism of Cr-based compounds.

The Cr-based compound $\CGN$ forms an antiperovskite-related structure. 
Its magnetic susceptibility has been reported  \cite{Nardin1, Lin}.
The structure of $\CGN$ at room temperature is tetragonal with the space group $P\bar{4}2_1m$ \cite{Boller1}
and transforms to another tetragonal structure with $I4/mcm$ at higher temperatures. 
In addition, there is a large hysteresis at the $P\bar{4}2_1m$ to $I4/mcm$ transition at temperatures in the range 300--400 K.
The susceptibility shows a Curie-Weiss behavior in the high temperature phase. Below the transition temperature $T^*$, the susceptibility exhibits a small value down to 5 K \cite{Lin}.
The previous magnetic susceptibility measurement suggested an antiferromagnetic state as the low temperature phase \cite{Lin}. However, there is a lack of microscopic information on the magnetic state of Cr. 

In the present study, we investigated the magnetic ground state of $\CGN$ using nuclear magnetic resonance (NMR) and  muon spin relaxation ($\muSR$) microscopy measurements, 
which reveal that Cr loses its moment below $T^*$, indicating that the magnetic ground state is not antiferromagnetic but nonmagnetic.

\section{Experimental}
A polycrystalline sample of $\CGN$ was prepared by a solid-state reaction using 
Cr$_2$N, Cr and Ge as the starting materials.
They were mixed with a stoichiometrical ratio, pressed into a pellet, sealed in an evacuated quartz tube and then calcined at 1173 K for 5 days.
The sample was characterized by room-temperature powder X-ray diffraction (XRD) measurement (X'Pert PRO Alpha-1, PANalytical). The sample primarily consisted of $\CGN$, although small amounts of impurities (1.5 \%  Cr$_2$O$_3$ and 1.1 \% GeO$_2$) were found, which do not substantially affect on our conclusion. 
The temperature dependence of magnetic susceptibility was measured at 2--1120 K using a SQUID magnetometer (MPMS, Quantum Design) and VSM by applying a magnetic field of up to 0.5 T.
NMR measurements were carried out using a phase-coherent spin-echo pulse spectrometer.
The $\Nitnuc$-NMR ($I = 1$, $\gamma /2 \pi = 3.075 \ \mathrm{MHz/ T}$) was measured in an external magnetic field of 6.19 T.
The NMR spectrum at 4.2 K was taken by summing the Fourier transform of the spin-echo signals at various frequencies.
The spin-lattice relaxation time was obtained by the saturation-recovery method.
Zero-field (ZF) $\mu$SR measurements were performed at the RIKEN-RAL Muon Facility at the Rutherford-Appleton Laboratory in the UK using a pulsed positive surface muon beam at 11.6--350 K.
 

\section{Results and Discussion} 
\begin{figure}[h]
	\centering
	\includegraphics[width=18pc, clip]{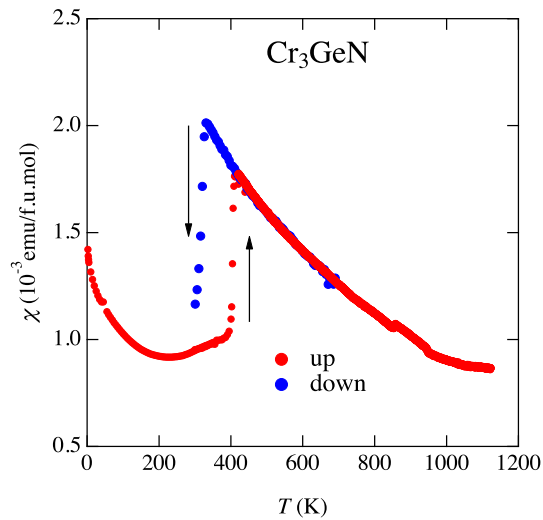}
	\caption{The temperature dependence of magnetic susceptibility for $\CGN$ at 2--1120 K.}
	\label{suscep}
\end{figure}

Figure \ref{suscep} shows the temperature dependence of the magnetic susceptibility. The result below 700 K corresponds well with data from the literature \cite{Nardin1, Lin}. The anomalies at 850 and 950 K correspond to other stractural transitions \cite{Wang}.

\begin{figure}[h]
\begin{minipage}[t]{16pc}
	\includegraphics[width=14pc, clip]{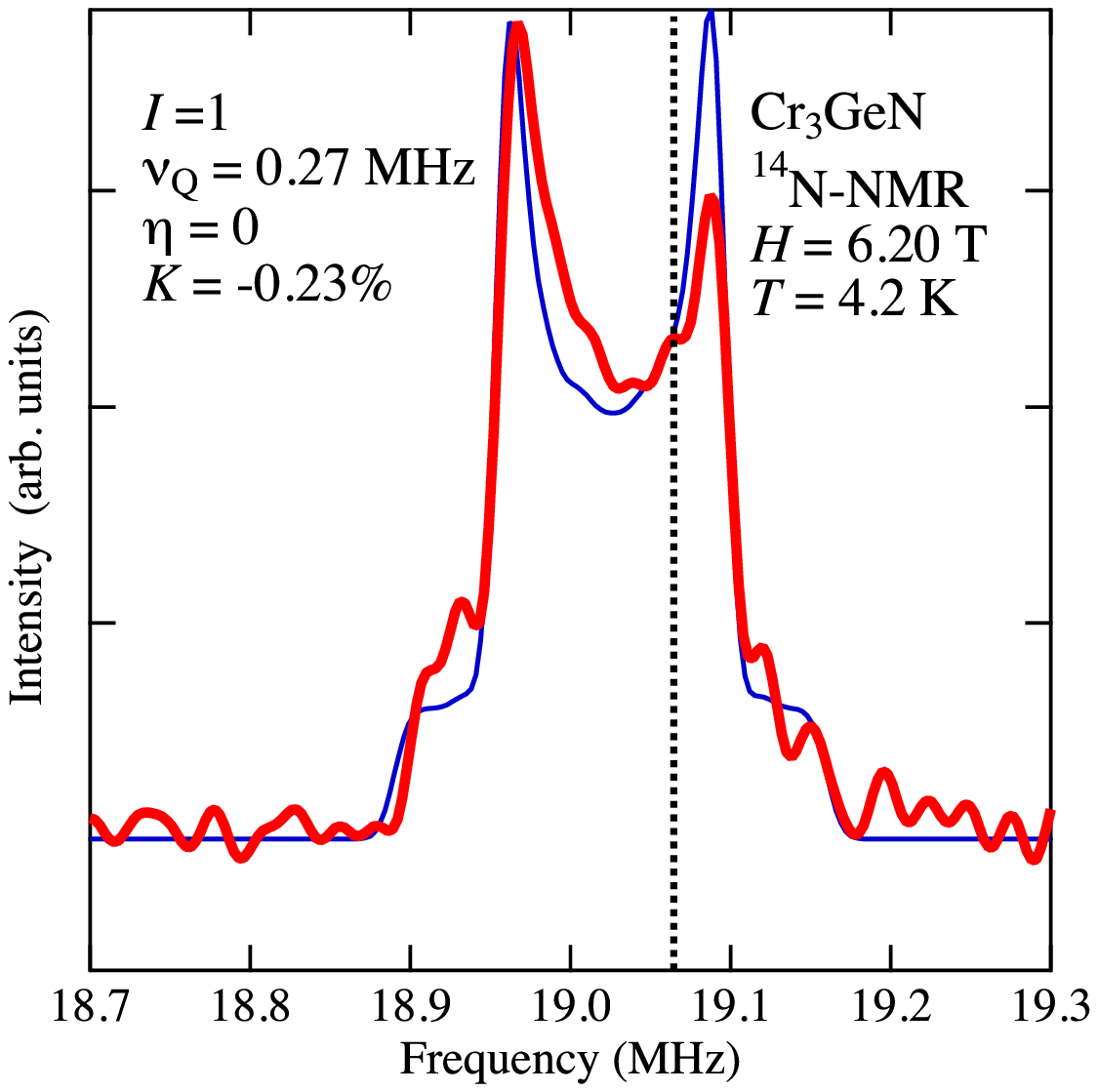}
	\caption{\label{NMR1} The Fourier transform $\Nitnuc$ NMR spectrum of $\CGN$ at 4.2 K. The red curve shows the experimental data, and blue curve is a simulation. The dotted line represents the frequency at $K = 0$.}
\end{minipage}\hspace{2pc}%
\begin{minipage}[t]{16pc}
	\includegraphics[width=15pc, clip]{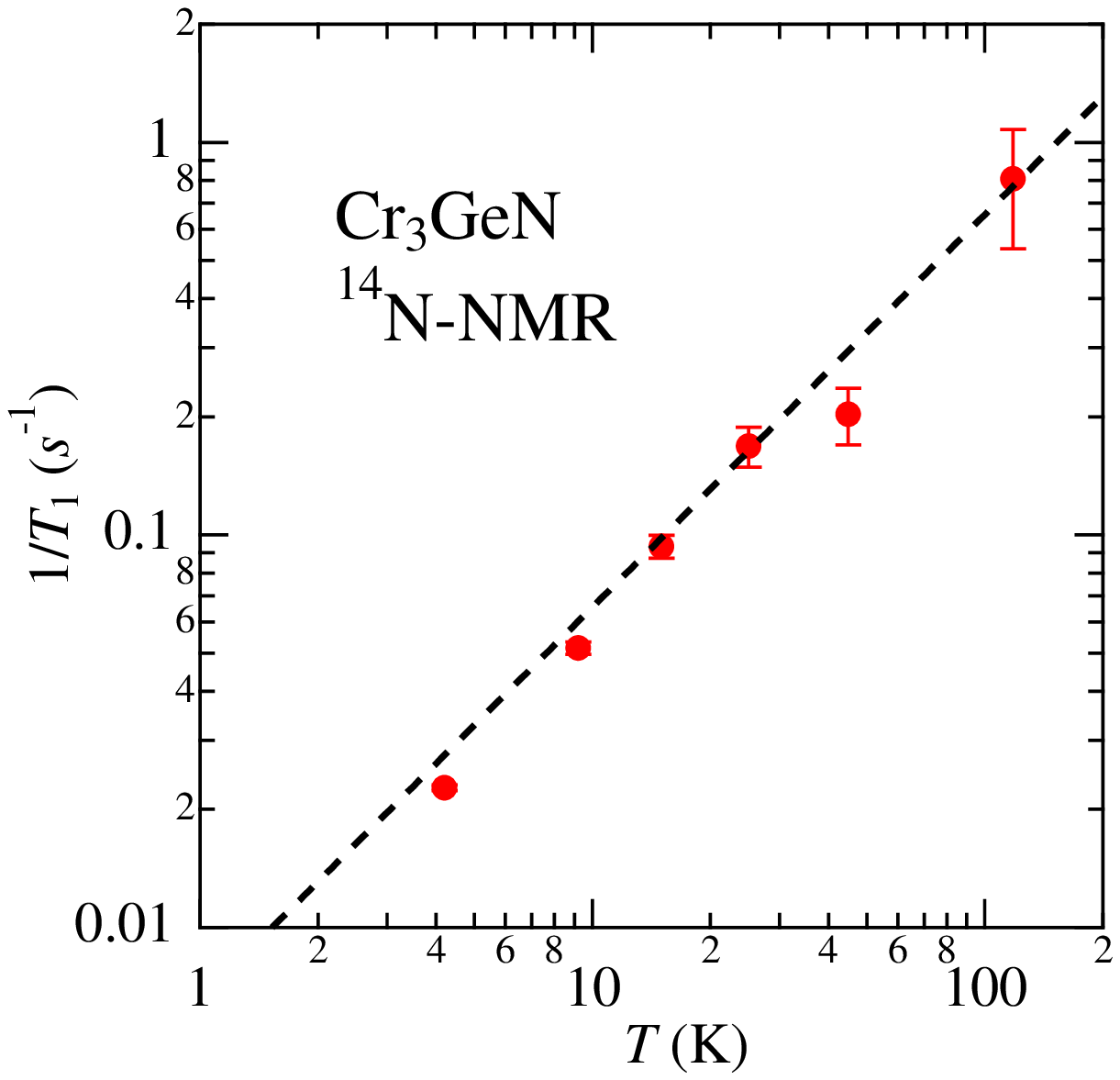}
	\caption{\label{NMR2} The temperature dependence of the $\Nitnuc$ spin-lattice relaxation rate for $\CGN$. The dashed line denotes the fit of experimental data with the Korringa law. }
\end{minipage}
\end{figure}

We performed microscopic measurements to demonstrate the disappearance of Cr moment in the ground state.
Figure \ref{NMR1} shows a $\Nitnuc$-NMR quadrupole-split spectrum.
The spectrum has a characteristic double peak structure and is well fitted by a simulated powder pattern for the $\Nitnuc$ nucleus with $I =1$ and the asymmetric parameter $\eta = 0$, since the nitrogen is located at an axially symmetric site. 
Both the quadrupole frequency, $\nu_Q$, and the shift, $K$, are small; $\nu_Q = 0.27$ MHz and $K = -0.23$\%. The small $K$ indicates no internal field in the ground state. 

$\Nitnuc$ spin-lattice relaxation measurements were performed at various temperatures up to 120 K. The spin-lattice relaxation time, $T_1$, was estimated by single exponential fitting. Figure \ref{NMR2} shows the temperature dependence of $1/ T_1$,  
which is proportional to $T$. 
The Korringa behavior indicates that the relaxation is dominated by conduction electrons. 
Moreover, $T_1$ is very long, say, 44 s at 4.2 K suggesting that the density of states at the Fermi level is markedly reduced at low temperature.
	
\begin{figure}[h]
	\centering
	\includegraphics[width=18pc, clip]{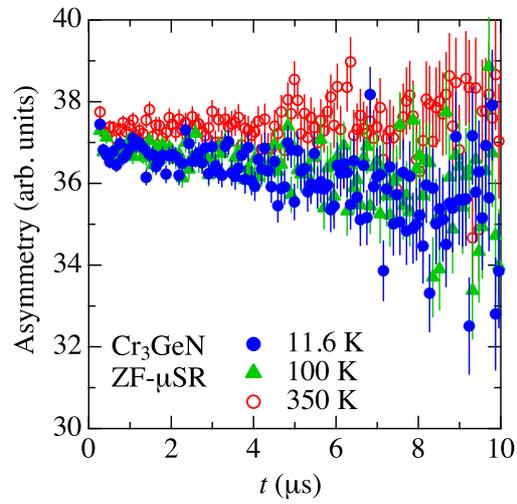}
	\caption{ZF-$\mu$SR spectra of $\CGN$ measured at 11.6, 100 and 350 K.}
	\label{muSR1}
\end{figure}

ZF-$\muSR$ measurements give useful information on the presence of the magnetic moment and magnetic ordering.
Figure \ref{muSR1} shows ZF-$\muSR$ spectra. 
At all temperatures, the muon spin asymmetry does not show any precession. This means that there is no magnetic ordering. 
In addition, the relaxation is very slow, which is consistent with the fact that the Cr magnetic moment disappears below $T^*$.

\section{Conclusion}
We investigated the magnetic ground state of $\CGN$ by using microscopic methods, $\Nitnuc$-NMR and $\muSR$, which demonstrated that the compound is the Pauli paramagnetic. 
ZF-$\muSR$ measurements suggest the disappearance of Cr moment below the structural transition, 
which is  different from the antiferromagnetism as expected.
The origin of the nonmagnetic ground state is still open to question. 
Band calculation would help reveal the mechanism of the disappearance of Cr moment at the structural transition.

\end{document}